\documentclass[aps,pre,reprint,superscriptaddress,footinbib]{revtex4-2}

\usepackage{amsmath}
\usepackage{amssymb}
\usepackage{amsfonts}
\usepackage{physics}
\usepackage{color}
\usepackage[colorlinks=true, linkcolor=blue, citecolor=blue, urlcolor=blue]{hyperref}
\hypersetup{
  pdftitle={Crossover of Scaling Behaviors of Work Cumulants in a Driven Gaussian Field Theory},
  pdfauthor={Yanbo Qiao, Ruohan Xu, and H. T. Quan}
}

\begin{document}
\title{Crossover of Scaling Behaviors of Work Cumulants in a Driven Gaussian Field Theory
}

\author{Yanbo Qiao}
\affiliation{School of Physics, Peking University, Beijing 100871, China}

\author{Ruohan Xu}
\affiliation{School of Physics, Peking University, Beijing 100871, China}

\author{H. T. Quan}
\email{htquan@pku.edu.cn}
\affiliation{School of Physics, Peking University, Beijing 100871, China}
\affiliation{Collaborative Innovation Center of Quantum Matter, Beijing 100871, China}
\affiliation{Frontiers Science Center for Nano-optoelectronics, Peking University, Beijing, 100871, China}

\date{\today}
\begin{abstract}
We derive the finite-temperature characteristic function of work (CFW)
for a driven $O(N)$ Gaussian field theory and obtain closed-form
expressions for all zero-temperature excess-work cumulants.
For gapped protocols, we use adiabatic perturbation theory (APT) to derive
the $\tau_Q^{-2}$ scaling for protocols with a nonzero first derivative at either boundary,
where $\tau_Q$ is the protocol duration; smoother boundaries lead to faster decay.
For power-law protocols approaching the critical point with exponent $p$,
we determine the competition between critical excitations and the
regular contribution.  The $n$th cumulant's critical contribution
follows Kibble--Zurek (KZ) scaling $\tau_Q^{-p(d+n)/(p+2)}$ for
$d+n<2p+4$, acquires a logarithmic correction $\tau_Q^{-2p}\log\tau_Q$
at $d+n=2p+4$, and scales as $\tau_Q^{-2p}$ above this condition,
where $d$ is the spatial dimension.
This analytical study of the APT--KZ crossover in a solvable model
provides a basis for studying their competition in interacting field theories.
\end{abstract}

\maketitle
\section{Introduction}
\label{sec:intro}
Work cumulants describe the energy transferred to a driven quantum system
and its fluctuations, providing information about nonequilibrium dynamics
beyond the average work.  In the two-point measurement scheme, they follow
from derivatives of the cumulant-generating function (CGF), the logarithm
of the characteristic function of work (CFW)
\cite{talknerFluctuationTheoremsWork2007,campisiColloquiumQuantumFluctuation2011,
espositoNonequilibriumFluctuationsFluctuation2009}.  Near a quantum critical
point, work cumulants can exhibit universal scaling associated with the
Kibble--Zurek (KZ) mechanism
\cite{Kibble1976,Zurek1985,feiWorkStatisticsQuantum2020,
feiUniversalScalingWork2021,zhangWorkStatisticsQuantum2022}.
Excitations produced at the protocol boundaries also contribute to work
fluctuations and are described by adiabatic perturbation theory (APT)
\cite{degrandiAdiabaticPerturbationTheory2010,
polkovnikovColloquiumNonequilibriumDynamics2011}.  Previous work derived
the KZ--APT crossover for linear protocols using independent quasiparticles
and a single-excitation approximation, with verification in the
transverse-field Ising chain \cite{feiWorkStatisticsQuantum2020}.
The KZ scaling of work cumulants under power-law protocols was also
derived using dimensional analysis in a conformal-field-theory framework
\cite{feiUniversalScalingWork2021}.
For finite-duration power-law protocols, a central question is whether
critical excitations govern the scaling of work cumulants or are outweighed
by regular adiabatic contributions.  Understanding this competition
requires determining both contributions and how the driving protocol
controls their crossover.

Field theory provides a general framework for phase transitions and their
universal dynamics.  The Gaussian field theory combines critical dynamics
with exact solvability through its decomposition into independent harmonic
oscillators with time-dependent frequencies.  It therefore allows us to
identify the roles of regular and critical excitations and explain their
crossover under power-law driving.  Path integrals
provide a representation of quantum evolution \cite{RevModPhys.20.367},
and the counting field for work can be implemented through a modified
real-time contour \cite{funoPathIntegralApproach2018}.  Qiu and one of us and collaborators
 evaluated this contour exactly for a harmonic oscillator
with an arbitrary time-dependent frequency
\cite{qiuPathIntegralApproach2020}.  The expression is equivalent to the
operator result of Deffner and Lutz
\cite{deffnerNonequilibriumWorkDistribution2008}.  At the many-body
level, extracting the scaling of work cumulants requires analyzing the
infrared and ultraviolet (UV) behavior of momentum integrals, which combine
the mode statistics with powers of the final excitation energy.

In this work, we study work statistics in a driven $O(N)$ Gaussian field
theory using the path-integral formulation and its independent oscillator
modes.  We derive the exact finite-temperature field CFW, verify the
Jarzynski equality, and obtain closed expressions for all zero-temperature
excess-work cumulants.  For gapped protocols, we derive the time-boundary contribution
 and show how the smoothness of the protocol at the initial and final moments determines the power-law
dependence of the cumulants on the driving duration.  For power-law
protocols ending at the critical point, we obtain the KZ scaling from the
rescaled mode equation and combine the critical contribution with the
regular contribution.  This gives the conditions under which
each contribution dominates the work cumulants.  We also determine when
the critical contribution follows KZ scaling, acquires a logarithmic
correction, or is dominated by its UV tail.

This article is organised as follows.
Section~\ref{sec:formalism_decoupling} introduces the field theory and its
decomposition into independent modes.
Section~\ref{sec:exact-CFW} derives the exact CFW.
Section~\ref{sec:exact-cumulants} obtains the zero-temperature excess-work
cumulants and analyzes their APT and KZ scaling and the crossover between
them.  Section~\ref{sec:conclusion} concludes the article.
The appendices give the calculation details.
\section{Field-Theoretic Formalism and Free-Field Decoupling}
\label{sec:formalism_decoupling}

We consider the driven $O(N)$ Gaussian field theory in a $d$-dimensional
periodic box of volume $V=L^d$.  In units with $\hbar=1$, the Lagrangian density for the
$N$-component real vector field
$\boldsymbol{\phi}(\mathbf{x},t)=(\phi_1,\ldots,\phi_N)$ is
\begin{equation}
    \mathcal{L}(t)
      =\frac{1}{2}(\partial_t\boldsymbol\phi)^2
       -\frac{c^2}{2}(\nabla\boldsymbol\phi)^2
       -\frac{r(t)}{2}\boldsymbol\phi^2.
\end{equation}
Here $c$ is the characteristic mode velocity, and $r(t)$ is a
time-dependent mass-squared parameter driving the system out of equilibrium.

We regularize the theory with a momentum cutoff $|\mathbf k|<\Lambda$,
which is kept fixed when varying the driving duration.  A spatial lattice
may be used instead, with the corresponding lattice dispersion and
momentum sums.

For a protocol of duration $\tau_Q$, we introduce the rescaled time
$s$ and write
\begin{equation}
\begin{aligned}
    r(t)&=\bar r(s),& s&=t/\tau_Q\in[0,1],\\
    r_i&=\bar r(0),& r_f&=\bar r(1).
\end{aligned}
    \label{eq:fixed-shape-protocol}
\end{equation}
The initial and final boundaries of the protocol are at $t=0$ and
$t=\tau_Q$, respectively.
In the slow-driving analysis, we vary $\tau_Q$ while keeping $\bar r(s)$ fixed.
A \emph{gapped protocol} obeys $\bar r(s)\ge r_{\min}>0$ for all $s$.  We
also consider the protocol ending at the critical point
\begin{equation}
    \bar r_p(s)=r_i(1-s)^p,\quad r_i>0,\quad p\in[1,\infty),\quad r_f=0.
    \label{eq:critical-protocol}
\end{equation}

The fluctuations of work are evaluated using the two-point measurement
scheme.  With
$H_i=H(0)$, $H_f=H(\tau_Q)$,
$U=\mathcal T\exp[-i\int_0^{\tau_Q}H(t)dt]$, and
$\rho_i=Z_i^{-1}e^{-\beta H_i}$, where $\mathcal T$ denotes time ordering,
$\beta=(k_BT)^{-1}$ is the inverse temperature, and
$Z_i=\operatorname{Tr}e^{-\beta H_i}$, the CFW reads~\cite{talknerFluctuationTheoremsWork2007}
\begin{equation}
    \chi_W(\lambda)
    =\operatorname{Tr}\!\left[
      e^{i\lambda H_f}Ue^{-i\lambda H_i}\rho_iU^\dagger
    \right].
    \label{eq:operator-CFW}
\end{equation}
The normalization $\chi_W(0)=1$ follows directly.  The Jarzynski
equality~\cite{jarzynskiNonequilibriumEqualityFree1997} is obtained from
$\chi_W(i\beta)$, as verified below.  Following the shifted
contour construction of Refs.~\cite{funoPathIntegralApproach2018,
qiuPathIntegralApproach2020,qiao2026field}, the many-body CFW is expressed as a
functional integral over a modified Schwinger--Keldysh contour.  The
counting field $\lambda$ introduces a relative temporal shift between the
forward ($+$) and backward ($-$) branches.  Inserting complete sets of
field eigenstates gives~\cite{qiao2026field}
\begin{equation}
\begin{aligned}
    \chi_W(\lambda)&=\frac{1}{Z_i}
      \int\mathcal D\boldsymbol\phi_+\mathcal D\boldsymbol\phi_-
      \mathcal D\boldsymbol\phi_E\\
    &\quad\times e^{iS_+[\boldsymbol\phi_+]-iS_-[\boldsymbol\phi_-]
      -S_E[\boldsymbol\phi_E]}.
\end{aligned}
\end{equation}
Here $S_E$ is the Euclidean action for the initial Hamiltonian $H_i$.
For real $\lambda$, the fields are continuous at the turning point,
$\boldsymbol{\phi}_{+}(\tau_Q+\lambda)
=\boldsymbol{\phi}_{-}(\tau_Q+\lambda)$, and satisfy the thermal sewing
conditions
$\boldsymbol{\phi}_+(0)=\boldsymbol{\phi}_E(\beta)$ and
$\boldsymbol{\phi}_-(0)=\boldsymbol{\phi}_E(0)$.  Complex counting fields, including
$\lambda=i\beta$, are reached by analytic continuation of the final CFW.

In the path-integral representation of the counting-field
insertions~\cite{funoPathIntegralApproach2018}, $S_+$ begins with evolution
for a duration $\lambda$ at $r_i$ and then follows the shifted protocol
$r(t-\lambda)$, while $S_-$ follows the unshifted protocol and is continued
for a duration $\lambda$ at $r_f$.  We now express these actions in
momentum space.

The discrete Fourier transform for each field component
$\alpha\in\{1,\ldots,N\}$ is
\begin{equation}
    \phi_\alpha(\mathbf x,t)
      =\frac{1}{\sqrt V}\sum_{\mathbf k}
       \phi_{\alpha,\mathbf k}(t)e^{i\mathbf k\cdot\mathbf x}.
\end{equation}
The reality condition $\phi_{\alpha,\mathbf{k}}^*
=\phi_{\alpha,-\mathbf{k}}$ leaves one complex degree of freedom for each
nonzero pair $\{\mathbf k,-\mathbf k\}$.  Let $\mathbb H_+^d$ contain one
representative of every such pair.  For $\mathbf{k}\in\mathbb H_+^d$ and
each component $\alpha$, write
$\phi_{\alpha,\mathbf k}(t)=[x_{\alpha,\mathbf k}(t)
+iy_{\alpha,\mathbf k}(t)]/\sqrt 2$, where
$x_{\alpha,\mathbf k}$ and $y_{\alpha,\mathbf k}$ are real.  Each nonzero
momentum pair therefore contributes two independent real oscillators per
field component.

Let $\widetilde{\chi}_{W,k}$, with $k=|\mathbf k|$, denote the CFW of one
such oscillator.  Factorization of the quadratic action gives, at finite
volume,
\begin{equation}
    \chi_W(\lambda)
    =\left[\widetilde{\chi}_{W,0}(\lambda)\right]^N
      \prod_{\mathbf k\in\mathbb H_+^d}
      \left[\widetilde{\chi}_{W,k}(\lambda)\right]^{2N},
    \label{eq:finite-mode-counting}
\end{equation}
Any additional self-conjugate lattice momenta contribute in the same way
as the zero mode.  At fixed cutoff, the thermodynamic-limit CGF density is
\begin{equation}
    \frac{1}{V}\log\chi_W(\lambda)
    =N\int_{|\mathbf k|<\Lambda}\frac{d^dk}{(2\pi)^d}
      \log\widetilde{\chi}_{W,k}(\lambda).
    \label{eq:thermodynamic-mode-counting}
\end{equation}
After mode decomposition, the Euclidean branch, together with the
corresponding normalization factor from $Z_i^{-1}$, gives the normalized thermal kernel
$\rho_k(x_i,x_i')$ for each real oscillator.  Its CFW has the contour
representation~\cite{funoPathIntegralApproach2018,qiuPathIntegralApproach2020}
\begin{equation}
\begin{split}
    \widetilde{\chi}_{W,k}(\lambda) = & \int dx_i dx_i' dx_f dx_f' \, \delta(x_f - x_f') \\
    & \times \int \mathcal{D}x \mathcal{D}x' \, e^{i S_{+,k}^\lambda[x] - i S_{-,k}^\lambda[x']} \rho_k(x_i, x_i'),
\end{split}
\end{equation}
Here $\omega_k^2(t)=c^2k^2+r(t)$.  The corresponding branch actions are
\begin{align}
    S_{+,k}^\lambda[x] =& \int_0^{\lambda} \left[ \frac{1}{2}\dot{x}^2 - \frac{1}{2}\omega_k^2(0)x^2 \right] dt \nonumber \\ 
    &+ \int_{\lambda}^{\tau_Q+\lambda} \left[ \frac{1}{2}\dot{x}^2 - \frac{1}{2}\omega_k^2(t - \lambda)x^2 \right] dt, \\
    S_{-,k}^\lambda[x'] =& \int_0^{\tau_Q} \left[ \frac{1}{2}\dot{x}'^2 - \frac{1}{2}\omega_k^2(t)x'^2 \right] dt \nonumber \\ 
    &+ \int_{\tau_Q}^{\tau_Q+\lambda} \left[ \frac{1}{2}\dot{x}'^2 - \frac{1}{2}\omega_k^2(\tau_Q)x'^2 \right] dt.
\end{align}
Together with Eqs.~\eqref{eq:finite-mode-counting} and
\eqref{eq:thermodynamic-mode-counting}, these expressions reduce the field
CFW to independent driven-oscillator problems labeled by $k$.  We evaluate
their common mode-resolved form in the next section.

\section{Exact Mode-Resolved Characteristic Function}
\label{sec:exact-CFW}
In this section, we assume positive initial and final frequencies for every
retained mode.

\subsection{Initial Thermal Density Matrix}
For an oscillator with initial frequency $\omega_k(0)$, this kernel is
given by~\cite{qiuPathIntegralApproach2020,
kleinertPathIntegralsQuantum2004},
\begin{equation}
\begin{split}
    \rho_k(x_i, x_i') =& \, 2\sinh\left(\frac{\beta\omega_k(0)}{2}\right) \left[ \frac{\omega_k(0)}{2\pi \sinh(\beta\omega_k(0))} \right]^{1/2} \\
    &\times \exp\Bigg\{ -\frac{\omega_k(0)}{2\sinh(\beta\omega_k(0))} \\
    &\quad \times \Big[ (x_i^2 + x_i'^2)\cosh(\beta\omega_k(0)) - 2x_i x_i' \Big] \Bigg\}.
\end{split}
\end{equation}
The prefactor fixes
$\operatorname{Tr}\rho_k=1$.

\subsection{Evaluation of the Path Integral}
\subsubsection{Exact quadratic propagator}
For a quadratic action with a time-dependent frequency $\omega_k(t)$,
write the path as the sum of its classical trajectory and a fluctuation,
\begin{equation}
    x(t) = x_\mathrm{cl}(t) + y(t).
\end{equation}
The classical path satisfies
$\ddot x_{\mathrm{cl}}+\omega_k^2(t)x_{\mathrm{cl}}=0$, with
$x_{\mathrm{cl}}(t_i)=x_i$ and $x_{\mathrm{cl}}(t_f)=x_f$, while
$y(t_i)=y(t_f)=0$.  The action separates as
\begin{equation}
    S[x] = S_\mathrm{cl}[x_\mathrm{cl}] + S_\mathrm{fluc}[y].
\end{equation}

The classical action is the boundary term
\begin{equation}
    S_\mathrm{cl}(x_f, t_f; x_i, t_i) = \frac{1}{2} \Big[ x_\mathrm{cl}(t) \dot{x}_\mathrm{cl}(t) \Big]_{t_i}^{t_f}.
\end{equation}
For piecewise evolution, continuity of $x_{\mathrm{cl}}$ and $\dot x_{\mathrm{cl}}$
cancels the contributions at each internal junction.

For a quadratic action, the fluctuation integral is Gaussian.  Evaluating
its determinant with the Gel'fand--Yaglom theorem gives the exact
propagator~\cite{gelfandIntegrationFunctionalSpaces1960,
kleinertPathIntegralsQuantum2004}:
\begin{equation}
    \int \mathcal{D}x \, e^{i S[x]} = \mathcal{F}_k(t_f, t_i) \, e^{i S_\mathrm{cl}(x_f, t_f; x_i, t_i)},
\end{equation}
where
\begin{equation}
    \mathcal{F}_k(t_f, t_i) = \sqrt{ \frac{i}{2\pi} \frac{\partial^2 S_\mathrm{cl}}{\partial x_i \partial x_f} }.
\end{equation}
We apply this result to the forward and backward actions,
$S_{+,k}^\lambda$ and $S_{-,k}^\lambda$.

\subsubsection{Backward-branch propagator}
The backward branch contains the driven evolution for $t\in[0,\tau_Q]$,
followed by constant-frequency evolution for
$t\in[\tau_Q,\tau_Q+\lambda]$.  The classical path obeys
$\ddot{x}_\mathrm{cl}+\omega_k^2(t)x_\mathrm{cl}=0$, with
$x_\mathrm{cl}(0)=x_i$ and $x_\mathrm{cl}(\tau_Q+\lambda)=x_f$.

For a fixed mode $k$, we write
$\omega_i\equiv\omega_{i,k}\equiv\omega_k(0)$ and
$\omega_f\equiv\omega_{f,k}\equiv\omega_k(\tau_Q)$.  Let $u(t)$ and $v(t)$
be two independent solutions during the driven segment, normalized by
$u(0)=1$, $\dot u(0)=0$, $v(0)=0$, and $\dot v(0)=1$.
Their Wronskian is
$W[u,v]=u\dot v-\dot uv=1$.

On the first segment,
\begin{equation}
    x_1(t) = x_i u(t) + \dot{x}_i v(t),
\end{equation}
where $\dot x_i\equiv\dot x_{\mathrm{cl}}(0)$ is fixed by the final boundary
condition.

For $t\geq\tau_Q$, the frequency is $\omega_f$.  Matching
$x_m=x_1(\tau_Q)$ and $\dot x_m=\dot x_1(\tau_Q)$ gives
\begin{equation}
    x_2(t) = x_m \cos\big[\omega_f (t-\tau_Q)\big] + \frac{\dot{x}_m}{\omega_f} \sin\big[\omega_f (t-\tau_Q)\big].
\end{equation}
With $\theta\equiv\omega_f\lambda$, the condition
$x_2(\tau_Q+\lambda)=x_f$ fixes
\begin{equation}
    \dot{x}_i = \frac{x_f - x_i \left[ u(\tau_Q) \cos\theta + \frac{\dot{u}(\tau_Q)}{\omega_f} \sin\theta \right]}{v(\tau_Q) \cos\theta + \frac{\dot{v}(\tau_Q)}{\omega_f} \sin\theta}.
\end{equation}

The classical action on the full backward branch is
\begin{equation}
    S_{-,\mathrm{cl}}^\lambda(x_i, x_f) = \frac{1}{2} \Big[ x_f \dot{x}_2(\tau_Q+\lambda) - x_i \dot{x}_1(0) \Big].
\end{equation}
In terms of the boundary coordinates, it has the quadratic form
\begin{equation}
    S_{-,\mathrm{cl}}^\lambda(x_i, x_f) = \frac{1}{2} \Big( C_{ii}^{(-)}x_i^2 + 2C_{if}^{(-)}x_i x_f + C_{ff}^{(-)}x_f^2 \Big).
\end{equation}
Define
\begin{equation}
    \mathcal{D}_- \equiv \omega_f v(\tau_Q) \cos\theta + \dot{v}(\tau_Q) \sin\theta.
\end{equation}
The Wronskian identity gives
\begin{align}
    C_{if}^{(-)} &= -\frac{\omega_f}{\mathcal{D}_-}, \\
    C_{ff}^{(-)} &= \frac{\omega_f \dot{v}(\tau_Q)\cos\theta - \omega_f^2 v(\tau_Q)\sin\theta}{\mathcal{D}_-}, \\
    C_{ii}^{(-)} &= \frac{\omega_f u(\tau_Q) \cos\theta + \dot{u}(\tau_Q) \sin\theta}{\mathcal{D}_-}.
\end{align}

The mixed derivative of the classical action is $C_{if}^{(-)}$, and the
fluctuation prefactor is
\begin{equation}
    \mathcal{F}_-(\lambda, \tau_Q) = \sqrt{ \frac{i}{2\pi} C_{if}^{(-)} }.
\end{equation}
Together with $S_{-,\mathrm{cl}}^\lambda$, this determines the
backward-branch propagator.

\subsubsection{Forward-branch propagator}
On the forward branch the temporal shift $\lambda$ delays the protocol.
For $t\in[0,\lambda]$, the frequency is fixed at $\omega_i$; for
$t\in[\lambda,\tau_Q+\lambda]$, it is $\omega_k(t-\lambda)$.

On $t\in[0,\lambda]$, let $\phi\equiv\omega_i\lambda$.  The
constant-frequency path joining $x(0)=x_i$ to $x(\lambda)=x_m$ is
\begin{equation}
    x_1(t) = \frac{1}{\sin\phi} \Big[ x_m \sin(\omega_i t) - x_i \sin\big(\omega_i (t-\lambda)\big) \Big].
\end{equation}
Its velocities at $t=0$ and $t=\lambda$ are
\begin{align}
    \dot{x}_1(0) &= \frac{\omega_i}{\sin\phi} (x_m - x_i \cos\phi), \label{eq:x1_dot_0} \\
    \dot{x}_1(\lambda) &= \frac{\omega_i}{\sin\phi} (x_m \cos\phi - x_i).
\end{align}

In the second segment $t \in [\lambda, \tau_Q+\lambda]$, introduce the
shifted time $\widetilde t=t-\lambda\in[0,\tau_Q]$.  The equation of motion
is $\ddot{x}_2+\omega_k^2(\widetilde t)x_2=0$, identical to that on the
driven segment of $S_-$.  The same basis $u(\widetilde t)$ and
$v(\widetilde t)$ therefore applies.  The classical path is parameterized by the
intermediate state $(x_m,\dot{x}_m)$ at $\widetilde t=0$:
\begin{equation}
    x_2(\widetilde t) = x_m u(\widetilde t) + \dot{x}_m v(\widetilde t).
\end{equation}
Velocity continuity, $\dot x_m=\dot x_1(\lambda)$, together with
$x_2(\tau_Q)=x_f$, determines the intermediate state.  Define
\begin{equation}
    \mathcal{D}_+ \equiv u(\tau_Q) \sin\phi + \omega_i v(\tau_Q) \cos\phi.
\end{equation}
Then
\begin{equation}
    x_m = \frac{x_f \sin\phi + x_i \omega_i v(\tau_Q)}{\mathcal{D}_+}.
\end{equation}

The forward classical action is
\begin{equation}
    S_{+,\mathrm{cl}}^\lambda(x_i, x_f) = \frac{1}{2} \Big[ x_f \dot{x}_2(\tau_Q) - x_i \dot{x}_1(0) \Big].
\end{equation}
Using Eq.~\eqref{eq:x1_dot_0} and the Wronskian identity gives
\begin{equation}
    S_{+,\mathrm{cl}}^\lambda(x_i, x_f) = \frac{1}{2} \Big( C_{ii}^{(+)}x_i^2 + 2C_{if}^{(+)}x_i x_f + C_{ff}^{(+)}x_f^2 \Big),
\end{equation}
where
\begin{align}
    C_{if}^{(+)} &= -\frac{\omega_i}{\mathcal{D}_+}, \\
    C_{ff}^{(+)} &= \frac{\dot{u}(\tau_Q)\sin\phi + \omega_i \dot{v}(\tau_Q)\cos\phi}{\mathcal{D}_+}, \\
    C_{ii}^{(+)} &= \frac{\omega_i u(\tau_Q)\cos\phi - \omega_i^2 v(\tau_Q)\sin\phi}{\mathcal{D}_+}.
\end{align}

The corresponding fluctuation prefactor is
\begin{equation}
    \mathcal{F}_+(\lambda, \tau_Q) = \sqrt{ \frac{i}{2\pi} \frac{\partial^2 S_{+,\mathrm{cl}}^\lambda}{\partial x_i \partial x_f} } = \sqrt{ \frac{i}{2\pi} C_{if}^{(+)} }.
\end{equation}
Together with $S_{+,\mathrm{cl}}^\lambda$, this determines the
forward-branch propagator.

\subsubsection{Single-mode CFW}
Combining the two branch propagators with the initial thermal kernel reduces
the CFW of each real oscillator to an integral over the temporal boundary
coordinates \cite{qiuPathIntegralApproach2020},
\begin{equation}
\begin{split}
    \widetilde{\chi}_{W,k}(\lambda) =& \int dx_i dx_i' dx_f \, \mathcal{F}_+(\lambda, \tau_Q) \mathcal{F}_-^*(\lambda, \tau_Q) \\
    &\times e^{i S_{+,\mathrm{cl}}^\lambda(x_i,x_f) - i S_{-,\mathrm{cl}}^\lambda(x_i',x_f)} \rho_k(x_i, x_i').
\end{split}
\end{equation}
For real $\lambda$, the backward evolution has the conjugated weight
$e^{-iS_{-,\mathrm{cl}}^\lambda}$ and prefactor $\mathcal F_-^*$, so that
$\mathcal F_+\mathcal F_-^*=(2\pi)^{-1}
\sqrt{C_{if}^{(+)}C_{if}^{(-)}}$.  The resulting analytic expression
defines the continuation to complex $\lambda$.

The remaining integral is Gaussian.  With
$\boldsymbol\xi=(x_i,x_i',x_f)^T$, write its exponent as
$-\boldsymbol\xi^T\mathbf M\boldsymbol\xi/2$.  Define
$K_\rho=\omega_i/[2\sinh(\beta\omega_i)]$,
$L_\rho=\cosh(\beta\omega_i)$,
and $\Delta C_{ff}=C_{ff}^{(+)}-C_{ff}^{(-)}$.  The quadratic form is
\begin{equation}
    \mathbf{M} = \begin{pmatrix}
        2K_{\rho}L_{\rho}-iC_{ii}^{(+)} & -2K_{\rho} & -iC_{if}^{(+)} \\
        -2K_{\rho} & 2K_{\rho}L_{\rho}+iC_{ii}^{(-)} & iC_{if}^{(-)} \\
        -iC_{if}^{(+)} & iC_{if}^{(-)} & -i\Delta C_{ff}
    \end{pmatrix}.
\end{equation}

The Gaussian integral gives
\begin{equation}
    \widetilde{\chi}_{W,k}(\lambda) = 2 \sinh\left(\frac{\beta \omega_i}{2}\right) \sqrt{\frac{2 K_\rho C_{if}^{(+)}C_{if}^{(-)}}{\det(\mathbf{M})}}.
\end{equation}
The field CFW follows from the regulated product
\eqref{eq:finite-mode-counting}, or equivalently from
Eq.~\eqref{eq:thermodynamic-mode-counting}.

\subsubsection{Determinant and generating function}

The identity $4K_\rho^2(L_\rho^2-1)=\omega_i^2$ simplifies the leading
$2\times2$ minor of $\mathbf M$.

Let $\mathcal{W} \equiv \det(\mathbf{M}) /
(2 K_\rho C_{if}^{(+)} C_{if}^{(-)})$.  Expanding the determinant and
dividing by the denominator in this definition gives
\begin{equation}
\begin{split}
    \mathcal{W} =& \, L_\rho \left( \frac{C_{if}^{(+)}}{C_{if}^{(-)}} + \frac{C_{if}^{(-)}}{C_{if}^{(+)}} \right) - 2 + \frac{L_\rho \Delta C_{ff} \Delta C_{ii}}{C_{if}^{(+)} C_{if}^{(-)}} \\
    &+ \frac{i}{2K_\rho} \left( \frac{C_{if}^{(+)} C_{ii}^{(-)}}{C_{if}^{(-)}} - \frac{C_{if}^{(-)} C_{ii}^{(+)}}{C_{if}^{(+)}} \right) \\
    &- \frac{i \Delta C_{ff}}{2K_\rho C_{if}^{(+)} C_{if}^{(-)}} \left( \omega_i^2 + C_{ii}^{(+)}C_{ii}^{(-)} \right),
\end{split}
\label{eq:W_expanded}
\end{equation}
where $\Delta C_{ii} = C_{ii}^{(-)} - C_{ii}^{(+)}$.

Substitution of the branch coefficients, followed by use of the Wronskian
identity, shows that the determinant depends on the protocol through the
two final-time combinations
$X=\omega_i^2(\dot v^2+\omega_f^2v^2)$ and
$Y=\dot u^2+\omega_f^2u^2$.  These combinations define the Husimi parameter
for this mode~\cite{husimiMiscellaneaElementaryQuantum1953}:
\begin{equation}
    Q^* = \frac{X + Y}{2\omega_i\omega_f}.
    \label{eq:Husimi-parameter}
\end{equation}
$Q^*\geq1$, with $Q^*=1$ when no excitations are produced.
The determinant reduction is given in Appendix~\ref{app:wronskian}.

The real part of the determinant is
\begin{equation}
    \mathcal{W}_{\mathrm{real}} = 2 \Big[ \cosh(\beta\omega_i) \big( \cos\phi\cos\theta + Q^* \sin\phi\sin\theta \big) - 1 \Big].
    \label{eq:W_real_final}
\end{equation}

The imaginary contribution is
\begin{equation}
    \mathcal{W}_{\mathrm{imag}} = 2 i \sinh(\beta\omega_i) \big( \sin\phi\cos\theta - Q^* \cos\phi\sin\theta \big).
    \label{eq:W_imag_final}
\end{equation}

Hence the single-mode CFW is
\begin{equation}
    \widetilde{\chi}_{W,k}(\lambda) = \frac{\sinh\left(\frac{\beta\omega_i}{2}\right)}{\sqrt{\Sigma_k(\lambda)}},
    \label{eq:chi_final}
\end{equation}
where $\Sigma_k(\lambda)\equiv\mathcal W(\lambda)/4$ is
\begin{equation}
\begin{split}
    \Sigma_k(\lambda) =& \, \frac{1}{2} \Big[ \cosh(\beta\omega_i) \big( \cos\phi\cos\theta + Q^* \sin\phi\sin\theta \big) - 1 \Big] \\
    &+ \frac{i}{2} \sinh(\beta\omega_i) \big( \sin\phi\cos\theta - Q^* \cos\phi\sin\theta \big).
\end{split}
    \label{eq:Sigma_final}
\end{equation}

The square-root branch in Eq.~\eqref{eq:chi_final} is fixed by analytic
continuation from $\lambda=0$.  Since
$\Sigma_k(0)=\sinh^2(\beta\omega_i/2)$, this gives
$\widetilde{\chi}_{W,k}(0)=1$.  Restoring the mode label where needed, all
dependence on the protocol enters through the Husimi parameter $Q_k^*$.
Equations
\eqref{eq:chi_final}--\eqref{eq:Sigma_final} are algebraically equivalent
to Eq.~(33) of Ref~\cite{qiuPathIntegralApproach2020} and
to the operator result of Deffner and Lutz
\cite{deffnerNonequilibriumWorkDistribution2008}.
\subsection{Verification of the Jarzynski Equality}
Setting $\lambda=i\beta$, the terms proportional to the Husimi parameter
cancel and
\begin{equation}
    \Sigma_k(i\beta) = \frac{1}{2} \big[ \cosh(\beta\omega_f) - 1 \big] = \sinh^2\left(\frac{\beta\omega_f}{2}\right).
\end{equation}
Equation~\eqref{eq:chi_final} then gives
\begin{equation}
    \widetilde{\chi}_{W,k}(i\beta) = \frac{\sinh\left(\frac{\beta\omega_i}{2}\right)}{\sinh\left(\frac{\beta\omega_f}{2}\right)} = \frac{Z_{k,f}}{Z_{k,i}} = e^{-\beta \Delta F_k},
\end{equation}
where $Z_{k,\mu}=[2\sinh(\beta\omega_{\mu,k}/2)]^{-1}$ for
$\mu=i,f$, and
$\Delta F_k\equiv-\beta^{-1}\log(Z_{k,f}/Z_{k,i})$.

With the full-momentum convention of
Eq.~\eqref{eq:thermodynamic-mode-counting}, the product over modes is
\begin{equation}
    \chi_W(i\beta) = \prod_{\mathbf{k}} \left[ e^{-\beta \Delta F_k} \right]^N = e^{-\beta \Delta F}.
\end{equation}
Here $\Delta F\equiv N\sum_{\mathbf k}\Delta F_k$ is the total free-energy
difference.  Thus the field CFW satisfies the quantum Jarzynski equality.

\section{Zero-Temperature Excess-Work Cumulants}
\label{sec:exact-cumulants}
At zero temperature, the adiabatic work is the ground-state energy difference
$W_{\mathrm{ad}}=E_{0,f}-E_{0,i}$.  Define the excess-work cumulant-generating
function density and its cumulants by
\begin{equation}
\begin{aligned}
    w_{\mathrm{ex}}(\lambda)
      &=\frac{1}{V}\left[\log\chi_W(\lambda)
        -i\lambda W_{\mathrm{ad}}\right],\\
    c_n
      &=\left.\partial_{(i\lambda)}^n
        w_{\mathrm{ex}}(\lambda)\right|_{\lambda=0}.
\end{aligned}
    \label{eq:excess-work-cumulants}
\end{equation}

Let $\omega_{i,k}\equiv\omega_k(0)$ and
$\omega_{f,k}\equiv\omega_k(\tau_Q)$.  To describe the excitations generated
by the protocol, we introduce Bogoliubov amplitudes $A_k(t)$ and $B_k(t)$.
With $\vartheta_k(t)=\int_0^t\omega_k(t')dt'$, they obey
\begin{equation}
\begin{aligned}
    \dot A_k(t)&=\frac{\dot\omega_k(t)}{2\omega_k(t)}
       e^{2i\vartheta_k(t)}B_k(t),\\
    \dot B_k(t)&=\frac{\dot\omega_k(t)}{2\omega_k(t)}
       e^{-2i\vartheta_k(t)}A_k(t),
\end{aligned}
\label{eq:Bogoliubov-equations}
\end{equation}
where a dot denotes $d/dt$, $A_k(0)=1$, and $B_k(0)=0$.  These equations
preserve $|A_k(t)|^2-|B_k(t)|^2=1$ and at $t=\tau_Q$ give
$Q_k^*=|A_k(\tau_Q)|^2+|B_k(\tau_Q)|^2=1+2|B_k(\tau_Q)|^2$.
Appendix~\ref{app:parameterization} derives this relation and connects the
Bogoliubov amplitudes to the complex and real mode bases.  The mean final occupation of a mode is
\begin{equation}
    b_k\equiv |B_k(\tau_Q)|^2=\frac{Q_k^*-1}{2}.
    \label{eq:b-definition}
\end{equation}
The final state evolved from the initial vacuum is a squeezed vacuum, so
excitations occur in pairs \cite{deffnerNonequilibriumWorkDistribution2008,
qiuPathIntegralApproach2020}.  The $\beta\to\infty$ limit of
Eqs.~\eqref{eq:chi_final} and
\eqref{eq:Sigma_final} is
\begin{equation}
    \widetilde{\chi}^{(0)}_{W,k}(\lambda)
      =\left[e^{i\lambda\omega_{i,k}}
       \left(\cos(\lambda\omega_{f,k})
       -iQ_k^*\sin(\lambda\omega_{f,k})\right)\right]^{-1/2}.
    \label{eq:zeroT-kernel-limit}
\end{equation}
Using $Q_k^*=1+2b_k$ gives the elementary identity
\begin{equation}
\begin{aligned}
    &e^{i\lambda\omega_{i,k}}
      \left[\cos(\lambda\omega_{f,k})
      -iQ_k^*\sin(\lambda\omega_{f,k})\right]\\
    &\qquad=e^{i\lambda(\omega_{i,k}-\omega_{f,k})}
       \left[1+b_k\left(1-e^{2i\lambda\omega_{f,k}}\right)\right].
\end{aligned}
\end{equation}
The branch fixed by $\widetilde{\chi}^{(0)}_{W,k}(0)=1$ therefore yields
\begin{equation}
    \widetilde{\chi}^{(0)}_{W,k}(\lambda)
    =e^{i\lambda(\omega_{f,k}-\omega_{i,k})/2}
      \left[1+b_k\bigl(1-e^{2i\lambda\omega_{f,k}}\bigr)
      \right]^{-1/2}.
    \label{eq:zeroT-CFW}
\end{equation}
After removing the deterministic adiabatic work, its logarithm is
\begin{align}
\widetilde w_{\mathrm{ex},k}(\lambda) = -\frac12\log\left[1+b_k\bigl(1-e^{2i\lambda\omega_{f,k}}\bigr)\right].
\label{eq:zeroT-CGF-series}
\end{align}
We define the corresponding mode cumulant by
$c_{n,k}=\left.\partial_{(i\lambda)}^n
\widetilde w_{\mathrm{ex},k}(\lambda)\right|_{\lambda=0}$.
For each regulated mode, $0\leq b_k/(1+b_k)<1$, and term-by-term
differentiation gives
\begin{equation}
    \left.\partial_{(i\lambda)}^n
      e^{2im\lambda\omega_{f,k}}\right|_{\lambda=0}
      =(2m\omega_{f,k})^n.
\end{equation}
Expanding Eq.~\eqref{eq:zeroT-CGF-series} either in $b_k/(1+b_k)$ or in
powers of $e^{2i\lambda\omega_{f,k}}-1$ gives the polylogarithm and
Stirling-number forms, respectively.  For every $n\in\mathbb N_+$,
\begin{equation}
    \begin{aligned}
    c_{n,k}
      &=2^{n-1}\omega_{f,k}^{n}
        \operatorname{Li}_{1-n}\!\left(\frac{b_k}{1+b_k}\right)\\
      &=\frac{(2\omega_{f,k})^n}{2}
        \sum_{\ell=1}^{n}(\ell-1)!S(n,\ell)b_k^\ell .
    \end{aligned}
    \label{eq:exact-mode-cumulants}
\end{equation}
Here $\operatorname{Li}_s(z)=\sum_{m\geq1}z^m/m^s$, and the Stirling
numbers of the second kind are defined by
$(e^x-1)^\ell=\ell!\sum_{n\geq\ell}S(n,\ell)x^n/n!$.  In particular,
\begin{equation}
    c_{n,k}
      =2^{n-1}\omega_{f,k}^{n}b_k+O(\omega_{f,k}^{n}b_k^2).
    \label{eq:mode-cumulant-small-b}
\end{equation}
For $b_k\ll1$, the term linear in $b_k$ gives the leading contribution
at every fixed cumulant order.
Using Eq.~\eqref{eq:thermodynamic-mode-counting}, the exact zero-temperature
field result is
\begin{equation}
    c_n
      =N2^{n-1}\int_{|\mathbf k|<\Lambda}
       \frac{d^dk}{(2\pi)^d}\,
       \omega_{f,k}^{n}\operatorname{Li}_{1-n}\!\left(
       \frac{b_k}{1+b_k}\right).
    \label{eq:exact-field-cumulants}
\end{equation}
Equation~\eqref{eq:exact-field-cumulants}, together with the Bogoliubov
equations~\eqref{eq:Bogoliubov-equations}, is the starting point for the
following analysis.  We first apply APT to gapped protocols, then examine
protocols ending at the critical point and derive their scaling behavior.
Section~\ref{sec:matching} combines these results and analyzes the crossover.

\subsection{Gapped Protocols: APT and the
\texorpdfstring{$\tau_Q^{-2}$}{tauQ-2} Law}
\label{sec:gapped-APT}

For a gapped protocol, we take $\tau_Q\to\infty$ at fixed $k$ in the exact
Bogoliubov equations and expand the mode amplitudes in inverse powers of
$\tau_Q$.  Assume that
$\bar r\in C^3[0,1]$, that $\bar r(s)\geq r_{\min}>0$, and that the cutoff
$\Lambda$ is fixed.  In the rescaled time $s=t/\tau_Q$, the mode frequency
and accumulated phase are
\begin{equation}
    \omega_k(s)=\sqrt{c^2k^2+\bar r(s)},\qquad
    \Theta_k(s)=\int_0^s\omega_k(\sigma)d\sigma .
\end{equation}
The phase in Eq.~\eqref{eq:Bogoliubov-equations} then satisfies
$\vartheta_k(\tau_Qs)=\tau_Q\Theta_k(s)$.
A prime in this subsection denotes $d/ds$.
To obtain the leading transition amplitude, we set $A_k=1$ in the
$B_k$ equation~\eqref{eq:Bogoliubov-equations} and integrate over time.
This is the first Born approximation \cite{degrandiAdiabaticPerturbationTheory2010}:
\begin{equation}
    B_k(\tau_Q)=\int_0^1ds\,
      \frac{\omega_k'(s)}{2\omega_k(s)}
      e^{-2i\tau_Q\Theta_k(s)}+O(\tau_Q^{-2}).
    \label{eq:B-first-APT}
\end{equation}
Since $\Theta_k'(s)=\omega_k(s)>0$, the phase has no stationary point on
$[0,1]$, and one integration by parts gives the APT expansion
\cite{degrandiAdiabaticPerturbationTheory2010}
\begin{align}
    B_k(\tau_Q)
    &=\frac{i}{4\tau_Q}\left[
      \frac{\omega_k'(1)}{\omega_{f,k}^2}
       e^{-2i\tau_Q\Theta_k(1)}
      -\frac{\omega_k'(0)}{\omega_{i,k}^2}\right]
      +O(\tau_Q^{-2})
      \nonumber\\
    &=\frac{i}{8\tau_Q}\mathcal E_k(\tau_Q)
      +O(\tau_Q^{-2}).
    \label{eq:B-endpoint}
\end{align}
Here the contributions from the two boundaries enter through
\begin{equation}
    \mathcal E_k(\tau_Q)
      \equiv\frac{\bar r'(1)}{\omega_{f,k}^3}
       e^{-2i\tau_Q\Theta_k(1)}
      -\frac{\bar r'(0)}{\omega_{i,k}^3}.
\label{eq:endpoint-combination}
\end{equation}
The gap $\omega_k(s)\geq\sqrt{r_{\min}}$ and the assumed smoothness
ensure that the $O(\tau_Q^{-2})$ remainder in Eq.~\eqref{eq:B-endpoint}
has a bound independent of $k$ at fixed cutoff $\Lambda$.
We therefore substitute this expansion into
Eq.~\eqref{eq:exact-mode-cumulants} and integrate over momentum to obtain
\begin{equation}
    \begin{aligned}
    c_n
      &=\frac{N2^{n-7}}{\tau_Q^2}
      \int_{|\mathbf k|<\Lambda}\frac{d^dk}{(2\pi)^d}
      \omega_{f,k}^{n}\left|\mathcal E_k(\tau_Q)\right|^2\\
      &\quad+O(\tau_Q^{-3}).
    \end{aligned}
    \label{eq:all-cumulant-APT}
\end{equation}
The $O(\tau_Q^{-3})$ remainder includes the interference between the first
and second adiabatic orders of $B_k(\tau_Q)$; the multipair terms in
Eq.~\eqref{eq:exact-mode-cumulants} begin only at $O(\tau_Q^{-4})$.

When the first derivatives at both boundaries are nonzero,
Eq.~\eqref{eq:all-cumulant-APT} contains an oscillatory cross term.  At
fixed finite volume, $\tau_Q^2c_n$ remains bounded but need not converge.
After taking the thermodynamic limit, the radial momentum integral of the
oscillatory cross term tends to zero as $\tau_Q\to\infty$, by the
Riemann--Lebesgue lemma\footnote{
The phase $\Theta_k(1)$ is strictly increasing for $k>0$.  Under
$u=\Theta_k(1)$, the radial weight near its lower integration limit is proportional
to $[u-\Theta_0(1)]^{(d-2)/2}$ and is integrable for $d>0$.}.
Thus, at fixed cutoff,
\begin{equation}
  \begin{aligned}
    c_n&=
      \frac{N2^{n-7}}{\tau_Q^2}
      \int_{|\mathbf k|<\Lambda}\frac{d^dk}{(2\pi)^d}
      \left\{[\bar r'(1)]^2\omega_{f,k}^{n-6}
      \right.\\
      &\hspace{5.5em}\left.
      +[\bar r'(0)]^2\frac{\omega_{f,k}^{n}}{\omega_{i,k}^{6}}\right\}
      +o(\tau_Q^{-2}).
  \end{aligned}
    \label{eq:dephased-endpoint-APT}
\end{equation}
If $\bar r'(0)=0$ and $\bar r'(1)\neq0$, only the first term
in Eq.~\eqref{eq:dephased-endpoint-APT} remains.
A boundary with a nonzero first derivative therefore contributes at order
$\tau_Q^{-2}$ to every fixed-order excess-work cumulant.
Appendix~\ref{app:spectral-representation} derives the final-boundary coefficient
in Eq.~\eqref{eq:dephased-endpoint-APT} from equilibrium correlation functions.
Appendix~\ref{app:repeated-ibp} extends the analysis to smoother boundaries.

For a boundary with a nonzero first derivative,
Eq.~\eqref{eq:B-endpoint} gives $b_k=O(\tau_Q^{-2}k^{-6})$ at large $k$.
The resulting APT coefficient is UV finite for $d+n<6$,
logarithmically cutoff dependent for $d+n=6$, and power-law cutoff
dependent for $d+n>6$.  For fixed $n$ and $\Lambda$, this affects only the
coefficient and does not modify the $\tau_Q^{-2}$ scaling.
\subsection{Critical-Ending Protocols: Kibble--Zurek Scaling}
\label{sec:KZ}
For the protocol in Eq.~\eqref{eq:critical-protocol}, the final
frequency $\omega_{f,k}=ck$ vanishes as $k\to0$.  The gap bound used
in Sec.~\ref{sec:gapped-APT} therefore no longer applies to the full
momentum integral.  We work with the thermodynamic-limit cumulant density
and rescale the mode equation to study these low-momentum modes\footnote{
At $r_f=0$, the isolated finite-volume zero mode is a zero-frequency
free particle and has no normalizable Fock vacuum.  We omit this single
mode when taking the thermodynamic limit.  Equivalently, one may keep
$r_f>0$, take the thermodynamic limit, and then send $r_f\to0^+$.}.
For $k>0$, we determine the same occupation $b_k$ that enters
Eq.~\eqref{eq:exact-field-cumulants} from the normalized complex mode
$z_k$, with $z_k(0)=1/\sqrt{\omega_{i,k}}$ and
$\dot z_k(0)=-i\sqrt{\omega_{i,k}}$.  For this normalization,
\begin{equation*}
    b_k=\frac{|\dot z_k(\tau_Q)|^2
      +\omega_{f,k}^2|z_k(\tau_Q)|^2}{4\omega_{f,k}}-\frac12,
\end{equation*}
where the dot denotes a physical-time derivative.
Set $t_r=\tau_Q-t$.  For the power-law protocol, this mode obeys
\begin{equation}
    \frac{d^2z_k}{dt_r^2}
      +\left[c^2k^2+r_i\left(\frac{t_r}{\tau_Q}\right)^p\right]z_k=0.
    \label{eq:critical-mode-equation}
\end{equation}
We choose the characteristic scales by imposing
$r_i(t_*/\tau_Q)^p=t_*^{-2}$ and $ck_*=t_*^{-1}$:
\begin{equation}
    t_*=\left(\frac{\tau_Q^p}{r_i}\right)^{1/(p+2)},
    \qquad k_*=\frac{1}{ct_*}
      \propto\tau_Q^{-p/(p+2)}.
    \label{eq:KZ-scales}
\end{equation}
For $ck\ll\sqrt{r_i}$, the maximum of the adiabatic parameter
$\epsilon_k(t)=|\dot\omega_k(t)|/\omega_k(t)^2$ satisfies
\begin{equation*}
    \max_{0\leq t\leq\tau_Q}\epsilon_k(t)
      \propto\left(\frac{k_*}{k}\right)^{(p+2)/p}.
\end{equation*}
Thus APT applies for $k\gg k_*$, but increasing $\tau_Q$ does not make
modes with $k\sim k_*$ adiabatic.  To follow these modes, we hold
$\kappa=k/k_*>0$ fixed and write
\begin{equation}
\begin{aligned}
    x&\equiv t_r/t_*,&
    \boldsymbol\kappa&\equiv\mathbf k/k_*,&
    \kappa&\equiv|\boldsymbol\kappa|=k/k_*,\\
    z_k(t)&=t_*^{1/2}f_{\kappa,x_i}(x),& x_i&=\tau_Q/t_*.
\end{aligned}
\end{equation}
Here $x_i$ is the initial scaled time; the final time is $x=0$.
The exact rescaled equation is
$f_{\kappa,x_i}''+\Omega_\kappa^2 f_{\kappa,x_i}=0$, with
$\Omega_\kappa(x)=\sqrt{\kappa^2+x^p}$ and primes denoting $d/dx$.
The initial vacuum gives
$f_{\kappa,x_i}(x_i)=\Omega_\kappa(x_i)^{-1/2}$ and
$f_{\kappa,x_i}'(x_i)=i\Omega_\kappa(x_i)^{1/2}$.
Thus the finite initial time remains in the boundary conditions.

Let $f_\kappa$ denote the exact solution selected by the
positive-frequency condition at infinity, independently of the finite
initial time $x_i$,
\begin{equation}
    f_\kappa(x)\sim\frac{1}{\sqrt{\Omega_\kappa(x)}}
      \exp\!\left[i\int^x\Omega_\kappa(x')dx'\right],
      \qquad x\to\infty.
\end{equation}
Since $f_\kappa$ and $f_\kappa^*$ form a basis of exact solutions,
\begin{equation*}
    f_{\kappa,x_i}(x)=\alpha_i f_\kappa(x)+\beta_i f_\kappa^*(x).
\end{equation*}
The coefficients $\alpha_i$ and $\beta_i$ satisfy
$|\alpha_i|^2-|\beta_i|^2=1$ and are fixed by the initial data at $x_i$.  At fixed $\kappa$, the large-$x_i$ WKB expansions of $f_\kappa$
and its derivative give
\begin{equation*}
    |\beta_i|\sim
    \frac{\Omega_\kappa'(x_i)}{4\Omega_\kappa(x_i)^2}
    \sim\frac{p}{8\sqrt{r_i}\,\tau_Q},
    \qquad \tau_Q\to\infty,
\end{equation*}
where $x_i^{p+2}=r_i\tau_Q^2$.  Evaluating the exact decomposition and
its derivative at $x=0$ expresses the final mode in terms of
$f_\kappa(0)$ and $f_\kappa'(0)$.  The occupation associated with $f_\kappa$ is
\begin{equation}
    \mathfrak b_p(\kappa)
      \equiv\frac{|f_\kappa'(0)|^2+\kappa^2|f_\kappa(0)|^2}{4\kappa}
      -\frac12.
    \label{eq:explicit-KZ-occupation}
\end{equation}
At fixed $\kappa>0$, these final mode values are finite.  Since
$|\alpha_i|^2=1+|\beta_i|^2$ and the overall mode phase does not affect
the occupation, the finite-$x_i$ corrections vanish as $\beta_i\to0$:
\begin{equation}
    b_{k_*\kappa}=\mathfrak b_p(\kappa)+o(1),
    \qquad \tau_Q\to\infty\ \text{at fixed }\kappa>0.
    \label{eq:occupation-scaling}
\end{equation}
For a given $p$, $\mathfrak b_p$ depends only on $\kappa$, so the leading
occupation profile depends on the protocol duration through $k/k_*$.  Under $\mathbf k=k_*\boldsymbol\kappa$, the exact cumulant
integral~\eqref{eq:exact-field-cumulants} becomes
\begin{equation}
\begin{aligned}
    c_n&=N2^{n-1}k_*^d(ck_*)^n
      \int_{|\boldsymbol\kappa|<\Lambda/k_*}
      \frac{d^d\kappa}{(2\pi)^d}\,\kappa^n\\
      &\qquad\times\operatorname{Li}_{1-n}\!\left(
      \frac{b_{k_*\kappa}}{1+b_{k_*\kappa}}\right).
\end{aligned}
    \label{eq:exact-rescaled-cumulants}
\end{equation}
The physical cutoff $\Lambda$ is fixed, so the rescaled upper limit
$\Lambda/k_*$ grows with $\tau_Q$.  The fixed-$\kappa$ limit in
Eq.~\eqref{eq:occupation-scaling} must therefore be supplemented by an
analysis of the large-$\kappa$ tail.  When the scaling moment converges,
we define the KZ contribution by
\begin{equation}
    c_n^{\mathrm{KZ}}
      =\mathcal A_{n,p}\,k_*^d(ck_*)^n
      \propto\tau_Q^{-p(d+n)/(p+2)}.
    \label{eq:KZ-cumulant-scaling}
\end{equation}
Although the region $k=O(k_*)$ shrinks, its occupation is not suppressed by
$\tau_Q$ at fixed $\kappa$; its phase-space and final-energy factors give
the power $k_*^{d+n}$ in Eq.~\eqref{eq:KZ-cumulant-scaling}.
The corresponding amplitude is
\begin{equation}
    \mathcal A_{n,p}=N2^{n-1}
      \int_{\mathbb R^d}\frac{d^d\kappa}{(2\pi)^d}\,\kappa^n
      \operatorname{Li}_{1-n}\!\left(
       \frac{\mathfrak b_p(\kappa)}
       {1+\mathfrak b_p(\kappa)}\right).
    \label{eq:KZ-amplitude}
\end{equation}
Section~\ref{sec:matching} establishes its convergence condition
$d+n<2p+4$ and treats the contribution from the initial boundary.
There is no infrared divergence of this moment for $d>0$:
the limiting mode and its derivative are finite at $\kappa=0$, so
$\mathfrak b_p(\kappa)=O(\kappa^{-1})$ and
Eq.~\eqref{eq:exact-mode-cumulants} gives
$\kappa^n\operatorname{Li}_{1-n}[\mathfrak b_p/(1+\mathfrak b_p)]=O(1)$.
The scaling in Eq.~\eqref{eq:KZ-cumulant-scaling} is consistent with the
KZ description of quantum critical dynamics \cite{dziarmagaQuantumIsingExact2005,
dziarmagaDynamicsQuantumPhase2010,zurekDynamicsQuantumPhase2005,
senDefectProductionNonlinear2008,polkovnikovUniversalAdiabaticDynamics2005}.
With dynamical exponent $z=1$ and correlation-length exponent $\nu=1/2$,
Eq.~\eqref{eq:KZ-cumulant-scaling} reproduces the general power-law work-cumulant
exponent $p(d+n z)\nu/(1+p z\nu)$
\cite{feiWorkStatisticsQuantum2020,feiUniversalScalingWork2021,
qiao2026field}.
\subsection{Crossover between KZ and APT scalings}
\label{sec:matching}

For the protocol in Eq.~\eqref{eq:critical-protocol}, we distinguish
excitations produced at the initial boundary from those associated with
the final boundary at criticality.  The latter include the low-momentum KZ region,
$k=O(k_*)$, and its adiabatic tail at $k\gg k_*$.  The two regions match
for $k_*\ll k\ll\sqrt{r_i}/c$.

At fixed $k>0$, the APT expansion of the Bogoliubov amplitude
separates the initial and final contributions,
$B_k=B_k^{\mathrm{init}}+B_k^{\mathrm{crit}}+\cdots$.
The initial boundary remains gapped and $\bar r_p'(0)=-p r_i$.
Its contribution from Eq.~\eqref{eq:B-endpoint} gives the regular APT term
\begin{equation}
    c_n^{\mathrm{init}}
      =\frac{N2^{n-7}p^2r_i^2}{\tau_Q^2}
       \int_{|\mathbf k|<\Lambda}\frac{d^dk}{(2\pi)^d}
       \frac{(ck)^n}{(c^2k^2+r_i)^3}.
    \label{eq:critical-initial-APT}
\end{equation}
The contribution of $k\lesssim k_*$ to this expression is
$O(\tau_Q^{-2}k_*^{d+n})=o(\tau_Q^{-2})$, so extending the regular
APT term to zero momentum does not change its leading coefficient.
We define $c_n^{\mathrm{crit}}$ by replacing $b_{k_*\kappa}$ with
$\mathfrak b_p(\kappa)$ in Eq.~\eqref{eq:exact-rescaled-cumulants}, retaining
the cutoff $\Lambda/k_*$.  This removes the finite-initial-time correction
and isolates the critical contribution, including both the
low-momentum region and its adiabatic tail.

To determine whether the KZ integral converges, we calculate its
large-$\kappa$ tail.  The adiabatic parameter
$|\Omega_\kappa'/\Omega_\kappa^2|$ becomes small throughout the evolution
as $\kappa\to\infty$, so the leading tail follows from the first Born
approximation to Eq.~\eqref{eq:Bogoliubov-equations} in the variables
$(x,\kappa)$.  Near the final boundary, where $x^p\ll\kappa^2$,
\begin{align}
    \frac{\Omega_\kappa'(x)}{2\Omega_\kappa(x)}
      &=\frac{p}{4\kappa^2}x^{p-1}
        +O(\kappa^{-4}x^{2p-1}),\\
    \int_0^x\Omega_\kappa(x')dx'
      &=\kappa x+O(\kappa^{-1}x^{p+1}).
    \label{eq:critical-endpoint-expansion}
\end{align}
Because the phase derivative $\Omega_\kappa(x)$ never vanishes, repeated
integration by parts suppresses the rapidly oscillating contribution away
from $x=0$.  With Abel regularization, the required integral is
\begin{equation}
    \lim_{\epsilon\downarrow0}\int_0^\infty dx\,x^{p-1}
      e^{(2i\kappa-\epsilon)x}
      =e^{i\pi p/2}\Gamma(p)(2\kappa)^{-p}.
    \label{eq:localized-endpoint-integral}
\end{equation}
Here $\Gamma$ is the Euler gamma function.  Multiplying this
integral by $p/(4\kappa^2)$ gives the magnitude of the leading
critical contribution to the amplitude
\begin{equation}
    |B_k^{\mathrm{crit}}(\tau_Q)|
      \sim\frac{\Gamma(p+1)}{2^{p+2}\kappa^{p+2}}
      =\frac{\Gamma(p+1)r_i}
       {2^{p+2}\tau_Q^p c^{p+2}k^{p+2}}.
    \label{eq:critical-endpoint-amplitude}
\end{equation}
The corresponding occupation has the tail
\begin{equation}
    \mathfrak b_p(\kappa)
      \sim\frac{\Gamma(p+1)^2}{2^{2p+4}}\kappa^{-2p-4}
      \qquad(\kappa\to\infty).
    \label{eq:critical-tail}
\end{equation}
Thus the large-$\kappa$ tail of the scaling function agrees with the
APT result at the final boundary.  The corresponding radial integral is
\begin{equation}
    \int_K^{\Lambda/k_*}d\kappa\,
      \kappa^{d+n-2p-5},\qquad K\gg1, 
    \label{eq:KZ-amplitude-UV}
\end{equation}
where $K$ is fixed as $\tau_Q\to\infty$.
Within $c_n^{\mathrm{crit}}$, the low-momentum contribution from
$0<k<Rk_*$, with $R$ fixed, follows KZ scaling
$\tau_Q^{-p(d+n)/(p+2)}$.  The large-$\kappa$ integral determines
whether the adiabatic tail changes this scaling.
Writing $D=d+n$ and $D_{\mathrm{UV}}=2p+4$, we obtain
\begin{equation}
    c_n^{\mathrm{crit}}\propto
    \begin{cases}
      k_*^D\propto\tau_Q^{-pD/(p+2)},&D<D_{\mathrm{UV}},\\
      k_*^{2p+4}\log(\Lambda/k_*)
        \propto\tau_Q^{-2p}\log\tau_Q,&D=D_{\mathrm{UV}},\\
      \Lambda^{D-2p-4}k_*^{2p+4}
        \propto\tau_Q^{-2p},&D>D_{\mathrm{UV}}.
    \end{cases}
    \label{eq:critical-scaling-regimes}
\end{equation}
For $D<D_{\mathrm{UV}}$, the scaling integral converges: the tail
contributes to the finite amplitude, and
$c_n^{\mathrm{crit}}\sim c_n^{\mathrm{KZ}}$ in
Eq.~\eqref{eq:KZ-cumulant-scaling}.
At $D=D_{\mathrm{UV}}$, the $\kappa^{-1}$ tail gives the logarithm.
For $D>D_{\mathrm{UV}}$, the tail determines the leading power and
its coefficient depends on the cutoff.  In the last two cases, the
low-momentum KZ contribution is subleading within $c_n^{\mathrm{crit}}$.

We next compare $c_n^{\mathrm{crit}}$ with the regular
contribution $c_n^{\mathrm{init}}\propto\tau_Q^{-2}$ in
Eq.~\eqref{eq:critical-initial-APT}.
Appendix~\ref{app:initial-correction} controls the remaining corrections,
including interference between the two boundaries, and establishes the leading result
in the thermodynamic limit:
\begin{equation}
    c_n\sim c_n^{\mathrm{init}}+c_n^{\mathrm{crit}}.
    \label{eq:combined-asymptotics}
\end{equation}
For $D<D_{\mathrm{UV}}$, the total cumulant follows KZ scaling when
$pD/(p+2)<2$ and the regular $\tau_Q^{-2}$ scaling when
$pD/(p+2)>2$.  For $p>1$, both contributions have the same power
at $D=2+4/p$.
At and above $D_{\mathrm{UV}}$, the regular contribution
dominates for $p>1$.  For $p=1$, the logarithm makes the critical
contribution dominant at $D=6$, while both boundaries contribute to
the leading $\tau_Q^{-2}$ coefficient for $D>6$.
Smoothing the initial boundary changes its decay power, as discussed
in Appendix~\ref{app:repeated-ibp}.

\section{Conclusion}
\label{sec:conclusion}

We have studied the scaling of work cumulants in a driven $O(N)$ Gaussian
field theory and the crossover between KZ scaling and APT.  We derived the exact
finite-temperature CFW and obtained closed expressions for the
zero-temperature excess-work cumulants.  This solvable model shows how
critical excitations and the smoothness of the protocol at the initial and final moments together determine work
fluctuations in quantum critical dynamics.  For gapped protocols, a
boundary with a nonzero first derivative gives a $\tau_Q^{-2}$ contribution
to every fixed-order cumulant, with higher powers for smoother boundaries.
For power-law protocols ending at the Gaussian critical point, the
critical contribution scales as $\tau_Q^{-p(d+n)/(p+2)}$ for
$d+n<2p+4$, acquires the logarithmic correction
$\tau_Q^{-2p}\log\tau_Q$ at $d+n=2p+4$, and is dominated by its
$\tau_Q^{-2p}$ adiabatic tail for $d+n>2p+4$.  Its competition with
the regular contribution determines the leading scaling
of the full cumulant.

These results provide further support for the universal scaling of work
cumulants and clarify the role of protocol boundaries in the crossover.
Our exact cumulants confirm the predicted KZ scaling
\cite{feiWorkStatisticsQuantum2020} beyond the single-excitation
approximation.  In the KZ regime, our results provide evidence supporting the
work-cumulant scaling derived in Ref.~\cite{qiao2026field}.

In future work, we will extend this analysis to interacting field theories
to study how interactions affect the scaling of work cumulants and the
crossover between critical excitations and adiabatic boundary contributions.

\begin{acknowledgments}
We acknowledge support from the National Natural Science Foundation of
China under Grants No.~12375028 and No.~12521004.
\end{acknowledgments}

\appendix
\section{Calculation of the Determinant}
\label{app:wronskian}

In Appendices~\ref{app:wronskian} and \ref{app:parameterization}, $k$ is
fixed and we write
$(u,v,z,A,B,Q^*)$ for
$(u_k,v_k,z_k,A_k,B_k,Q_k^*)$, together with
$\omega_i\equiv\omega_{i,k}$ and $\omega_f\equiv\omega_{f,k}$.  They also
write $\omega(t)$ and $\vartheta(t)$ for $\omega_k(t)$ and
$\vartheta_k(t)$.

We simplify the path-integral determinant $\mathcal W$ using
$(u,\dot u,v,\dot v)$ for the basis functions and their derivatives at
$t=\tau_Q$.  Define
$\Delta C_{ff}=P/(\mathcal D_+\mathcal D_-)$ and
$\Delta C_{ii}=R/(\mathcal D_+\mathcal D_-)$.  Their numerators are
\begin{equation}
    P=-\omega_f\sin\phi\cos\theta
      +\frac{X}{\omega_i}\cos\phi\sin\theta
      +P_3\sin\phi\sin\theta,
\end{equation}
where $X=\omega_i^2(\dot v^2+\omega_f^2v^2)$ and
$P_3=\dot u\dot v+\omega_f^2uv$, and
\begin{equation}
    R=R_1\sin\phi\cos\theta-\omega_i\cos\phi\sin\theta
      +R_3\sin\phi\sin\theta,
\end{equation}
where $R_1=\omega_f(u^2+\omega_i^2v^2)$ and
$R_3=u\dot u+\omega_i^2v\dot v$.

For the real part, define
$S\equiv\omega_i^2\mathcal D_-^2+\omega_f^2\mathcal D_+^2+PR$ and write
$S=S_{cc}\cos^2\phi+S_{ss}\sin^2\phi
+S_{cs}\sin\phi\cos\phi$.  The Wronskian identity
$u\dot v-\dot uv=1$ gives
\begin{align}
    S_{cc}&=2\omega_i\omega_f(v\cos\theta)
      (\omega_i\mathcal D_-),\\
    S_{ss}&=2\omega_i\omega_fQ^*(u\sin\theta)\mathcal D_-,\\
    S_{cs}&=2\omega_i\omega_f
      \bigl(u\cos\theta+\omega_iQ^*v\sin\theta\bigr)\mathcal D_-,
\end{align}
where $Y\equiv\dot u^2+\omega_f^2u^2$ and
$Q^*=(X+Y)/(2\omega_i\omega_f)$.  Therefore
\begin{equation}
    S = 2\omega_i\omega_f \mathcal{D}_+ \mathcal{D}_- \big( \cos\phi\cos\theta + Q^* \sin\phi\sin\theta \big),
\end{equation}
which gives Eq.~\eqref{eq:W_real_final}.

For the imaginary component, define
$N_+\equiv\mathcal D_+C_{ii}^{(+)}$ and
$N_-\equiv\mathcal D_-C_{ii}^{(-)}$.  The reduced numerator is
$S_I\equiv2K_\rho\omega_i\omega_f\mathcal D_+\mathcal D_-
\mathcal W_{\mathrm{imag}}/i=J_1-J_2$, where
$J_1=\omega_i^2\mathcal D_-N_- -\omega_f^2\mathcal D_+N_+$ and
$J_2=P(\omega_i^2\mathcal D_+\mathcal D_-+N_+N_-)$.  Direct expansion
shows that $S_I$ is divisible by $\mathcal D_+\mathcal D_-$; define
$Q_I=S_I/(\mathcal D_+\mathcal D_-)$.

Expand
$Q_I=C_{cc}\cos\phi\cos\theta+C_{ss}\sin\phi\sin\theta
+C_{sc}\sin\phi\cos\theta+C_{cs}\cos\phi\sin\theta$.
Projection onto this harmonic basis and use of the Wronskian identity give
\begin{equation}
    C_{cc}=C_{ss}=0,\qquad
    C_{sc}=2\omega_i^2\omega_f,\qquad
    C_{cs}=-2\omega_i^2\omega_fQ^*.
\end{equation}
Substitution into $Q_I$ gives Eq.~\eqref{eq:W_imag_final}.

\section{Parameterization of the Mode Dynamics}
\label{app:parameterization}
The normalized complex mode is related to the real basis functions by
\begin{equation}
    z(t)=\frac{u(t)}{\sqrt{\omega_i}}
      -i\sqrt{\omega_i}\,v(t).
\end{equation}
To transform $\ddot{z}+\omega^2(t)z=0$ into coupled first-order equations,
use the physical-time phase $\vartheta(t)$ defined in the main text and the
ansatz~\cite{lewisExactQuantumTheory1969}
\begin{equation}
    z(t)=\frac{1}{\sqrt{\omega(t)}}\left[
      A(t)e^{-i\vartheta(t)}+B(t)e^{i\vartheta(t)}\right]
\end{equation}
and impose the standard constraint
\begin{equation}
    \dot{z}(t) = -i\sqrt{\omega(t)} \Big[
      A(t)e^{-i\vartheta(t)}-B(t)e^{i\vartheta(t)} \Big].
\end{equation}
Substitution into the mode equation gives
\begin{equation}
    \dot{A}=\frac{\dot{\omega}}{2\omega}e^{2i\vartheta(t)}B,
    \qquad
    \dot{B}=\frac{\dot{\omega}}{2\omega}e^{-2i\vartheta(t)}A.
    \label{eq:appendix-Bogoliubov-equations}
\end{equation}
The initial conditions $z(0)=1/\sqrt{\omega_i}$ and
$\dot z(0)=-i\sqrt{\omega_i}$ imply $A(0)=1$ and $B(0)=0$.  The equations
preserve $|A(t)|^2-|B(t)|^2=1$.

Expressing the Husimi parameter $Q^*$ in terms of the final state gives
\begin{equation}
    Q^* = \frac{1}{2\omega_f}
      \left[\omega_f^2|z(\tau_Q)|^2+|\dot z(\tau_Q)|^2\right].
\end{equation}
Substituting the Bogoliubov parameterization, the cross terms cancel and
give $Q^*=|A(\tau_Q)|^2+|B(\tau_Q)|^2$.  Together with
$|A(\tau_Q)|^2=1+|B(\tau_Q)|^2$, this yields, after restoring the mode label,
$Q_k^*-1=2|B_k(\tau_Q)|^2$.

\section{Analysis of the smoothness of the protocol at the initial and final moments and Spectral Representation}
\label{app:endpoint-details}

\subsection{Repeated integration by parts}
\label{app:repeated-ibp}

For a gapped protocol, if the first nonzero derivative at a boundary is
of integer order $q\geq1$ and $\bar r\in C^{q+2}[0,1]$, its amplitude is
$O(\tau_Q^{-q})$ and its contribution to each fixed-order cumulant is
$O(\tau_Q^{-2q})$.
The boundary with the lowest nonvanishing derivative order determines
the leading power of $\tau_Q^{-1}$.

Under the uniform-gap and fixed-cutoff assumptions of
Sec.~\ref{sec:gapped-APT}, $\Theta_k'(s)=\omega_k(s)>0$.  To organize the
APT expansion, define
\begin{equation}
    f_{0,k}(s)=\frac{\omega_k'(s)}{2\omega_k(s)},
    \qquad
    f_{j+1,k}(s)=\frac{d}{ds}
      \left[\frac{f_{j,k}(s)}{\omega_k(s)}\right].
\end{equation}
The first Born approximation to $B_k(\tau_Q)$ is then
\[
    B_k^{(1)}(\tau_Q)=\int_0^1ds\,f_{0,k}(s)e^{-2i\tau_Q\Theta_k(s)}.
\]
For $\bar r\in C^{q+2}[0,1]$, repeated integration by parts gives
\begin{equation}
\begin{aligned}
    B_k^{(1)}(\tau_Q)&=\sum_{j=0}^{q-1}(-1)^j
      \left(\frac{i}{2\tau_Q}\right)^{j+1}\\
      &\quad\times\left[
       \frac{f_{j,k}(s)}{\omega_k(s)}
       e^{-2i\tau_Q\Theta_k(s)}\right]_{0}^{1}
      +O(\tau_Q^{-q-1}).
\end{aligned}
    \label{eq:repeated-endpoint-IBP}
\end{equation}
If both boundaries are flat through order $q-1$, expanding the coupled
Bogoliubov equations in Eq.~\eqref{eq:appendix-Bogoliubov-equations} gives
\begin{equation}
\begin{aligned}
    B_k(\tau_Q)
      &=-\frac{(-i)^q}{2^{q+2}\tau_Q^q}
        \left[\frac{\bar r^{(q)}(s)}{\omega_k(s)^{q+2}}
        e^{-2i\tau_Q\Theta_k(s)}\right]_0^1\\
      &\quad+O(\tau_Q^{-q-1}).
\end{aligned}
    \label{eq:exact-smoothed-endpoint}
\end{equation}
The leading coefficient agrees with Eq.~\eqref{eq:repeated-endpoint-IBP};
feedback between $A_k$ and $B_k$ affects only higher-order terms in this
APT expansion.
Taking $q$ to be the lowest nonvanishing derivative order at either
boundary gives $c_n=O(\tau_Q^{-2q})$ for every fixed $n$.
When both boundaries contribute at order $\tau_Q^{-q}$, their amplitudes
produce an oscillatory cross term in $|B_k(\tau_Q)|^2$.

For a protocol ending at criticality, one may instead smooth only the
initial boundary while retaining the same power-law approach with exponent
$p$ at the final boundary.  If the first nonzero initial derivative is of
order $q$, the regular initial contribution scales as $\tau_Q^{-2q}$.
For $D<D_{\mathrm{UV}}$, it competes with $\tau_Q^{-pD/(p+2)}$;
for $D=D_{\mathrm{UV}}$, with $\tau_Q^{-2p}\log\tau_Q$; and for
$D>D_{\mathrm{UV}}$, with $\tau_Q^{-2p}$.
The slower-decaying contribution dominates.  At the marginal condition,
the logarithm makes the critical term dominant when $p=q$; above it, the
two boundaries are of the same order when $p=q$, and their cross term must
be considered when determining the coefficient.

\subsection{Finite-initial-time correction}
\label{app:initial-correction}

Write $b_c=\mathfrak b_p(k/k_*)$ and $a_k=|\beta_i|^2$.
The regular expansion gives, uniformly for $0\leq k<\Lambda$,
\begin{equation}
    a_k=\frac{p^2r_i^2}{64\tau_Q^2\omega_{i,k}^6}
      +O(\tau_Q^{-3}).
\end{equation}
The exact mode decomposition in Sec.~\ref{sec:KZ} gives
\begin{equation}
\begin{aligned}
    b_k&=b_c+a_k(1+2b_c)+\delta b_k,\\
    |\delta b_k|&\leq2\sqrt{a_k(1+a_k)b_c(1+b_c)}.
\end{aligned}
\end{equation}
The term linear in $a_k$ gives $c_n^{\mathrm{init}}$.
The polynomial cumulants in Eq.~\eqref{eq:exact-mode-cumulants} and the
small- and large-$\kappa$ bounds on $b_c$ then yield
\begin{equation}
\begin{aligned}
  |c_n-c_n^{\mathrm{init}}-c_n^{\mathrm{crit}}|
  &\lesssim \tau_Q^{-3}+\tau_Q^{-1}k_*^D\\
  &\quad\times\left[1+\int_1^{\Lambda/k_*}
    d\kappa\,\kappa^{D-p-3}\right].
\end{aligned}
\end{equation}
At fixed cutoff and cumulant order, this is subleading for $p>1$
and for $p=1$, $D\leq6$.
For $p=1$, $D>6$, the remaining order-$\tau_Q^{-2}$ cross term has
radial weight proportional to $k^{D-4}/\omega_{i,k}^3$ and phase
$2\tau_Q\Theta_k(1)$.  On $k\geq\delta>0$, the APT expansion
applies and the Riemann--Lebesgue lemma suppresses its integral.
The bound above controls the $k<\delta$ part by
$O(\tau_Q^{-2}\delta^{D-3})+o(\tau_Q^{-2})$ at fixed $\delta$.
Taking $\tau_Q\to\infty$ and then $\delta\to0$ gives
Eq.~\eqref{eq:combined-asymptotics}.

\subsection{Many-body spectral representation}
\label{app:spectral-representation}

For a gapped zero-temperature protocol with $\bar r'(0)=0$, the leading
final-boundary coefficient has a spectral representation.  We first work
at fixed volume and regulator and denote fixed-volume remainders by
$O_V(\cdot)$.
Let $|0\rangle$ and $|m\rangle$ be final eigenstates, with
$\Omega_m=E_m-E_0>0$, and define
$\mathcal X_f=\left.\partial_rH(r)\right|_{r=r_f}$ and
\begin{equation}
    \mathcal S_{\mathcal X}^{(0)}(\Omega)
      =\frac{1}{V}\sum_{m\ne0}
       |\langle m|\mathcal X_f|0\rangle|^2
       \delta(\Omega-\Omega_m).
    \label{eq:generalized-force-spectrum}
\end{equation}
Differentiating the ground-state eigenvalue equation gives
$\langle m|\partial_r0\rangle
=-\langle m|\mathcal X_f|0\rangle/\Omega_m$; integration by parts
supplies another inverse gap
\cite{degrandiAdiabaticPerturbationTheory2010,
polkovnikovUniversalAdiabaticDynamics2005}, yielding
\begin{equation}
    P_m^{(f)}
      =\frac{[\bar r'(1)]^2}{\tau_Q^2}
       \frac{|\langle m|\mathcal X_f|0\rangle|^2}{\Omega_m^4}
       +O_V(\tau_Q^{-3}).
    \label{eq:spectral-transition-probability}
\end{equation}
Since $P_m^{(f)}=O_V(\tau_Q^{-2})$, terms quadratic in the excitation
probabilities enter the cumulants only at $O_V(\tau_Q^{-4})$.  Thus, for
each fixed $n$,
\begin{equation}
    c_n^{(f)}
      =\frac{[\bar r'(1)]^2}{\tau_Q^2}
       \int_0^\infty d\Omega\,
       \Omega^{n-4}\mathcal S_{\mathcal X}^{(0)}(\Omega)
       +O_V(\tau_Q^{-3}).
\end{equation}
For $n=2$, the spectral moment is the ground-state quantum metric
associated with $r$; its relation to the leading work variance for a
linear protocol was derived in Ref.~\cite{kolodrubetzClassifyingMeasuringGeometry2013}.

For the Gaussian field,
$\mathcal X_f=\frac12\int d^dx\,\boldsymbol\phi^2$.
Using
$\phi_{\alpha,\mathbf k}
=(a_{\alpha,\mathbf k}+a_{\alpha,-\mathbf k}^{\dagger})/
\sqrt{2\omega_{f,k}}$, each unordered pair
$\{\mathbf k,-\mathbf k\}$ has squared matrix element
$1/(4\omega_{f,k}^2)$; a self-conjugate mode contributes
$1/(8\omega_{f,k}^2)$.  The full-momentum convention therefore gives
\begin{equation}
    \mathcal S_{\mathcal X}^{(0)}(\Omega)
      =\frac{N}{8}\int_{|\mathbf k|<\Lambda}
       \frac{d^dk}{(2\pi)^d}
       \frac{\delta(\Omega-2\omega_{f,k})}{\omega_{f,k}^2}.
    \label{eq:phi2-spectral-density}
\end{equation}
At finite volume, the integral is replaced by
$V^{-1}\sum_{\mathbf k}$, including the self-conjugate modes.  For the
regulated Gaussian theory, the exact mode CGF controls the remainder at
every fixed cumulant order and the thermodynamic limit at fixed regulator.
Substituting Eq.~\eqref{eq:phi2-spectral-density} into the spectral moment
above reproduces the final-boundary term in
Eq.~\eqref{eq:dephased-endpoint-APT}.  The leading
final-boundary contribution has the same spectral form for a general gapped
Hamiltonian with a nondegenerate ground state, provided the APT
expansion is uniform and the weighted spectral integral is finite.

\vspace{12pt}
\bibliographystyle{apsrev4-2}
\bibliography{KZM_Work_Statistics}

@article{campisiColloquiumQuantumFluctuation2011,
  title = {{\emph{Colloquium}} : {{Quantum}} Fluctuation Relations: {{Foundations}} and Applications},
  shorttitle = {{\emph{Colloquium}}},
  author = {Campisi, Michele and H{\"a}nggi, Peter and Talkner, Peter},
  year = 2011,
  month = jul,
  journal = {Reviews of Modern Physics},
  volume = {83},
  number = {3},
  pages = {771--791},
  issn = {0034-6861, 1539-0756},
  doi = {10.1103/RevModPhys.83.771},
  urldate = {2026-05-29},
  copyright = {http://link.aps.org/licenses/aps-default-license},
  langid = {english}
}

@article{deffnerNonequilibriumWorkDistribution2008,
  title = {Nonequilibrium Work Distribution of a Quantum Harmonic Oscillator},
  author = {Deffner, Sebastian and Lutz, Eric},
  year = 2008,
  month = feb,
  journal = {Physical Review E},
  volume = {77},
  number = {2},
  pages = {021128},
  issn = {1539-3755, 1550-2376},
  doi = {10.1103/PhysRevE.77.021128},
  urldate = {2026-02-15},
  copyright = {http://link.aps.org/licenses/aps-default-license},
  langid = {english}
}

@incollection{degrandiAdiabaticPerturbationTheory2010,
  title = {Adiabatic {{Perturbation Theory}}: {{From Landau}}--{{Zener Problem}} to {{Quenching Through}} a {{Quantum Critical Point}}},
  shorttitle = {Adiabatic {{Perturbation Theory}}},
  booktitle = {Quantum {{Quenching}}, {{Annealing}} and {{Computation}}},
  author = {De Grandi, C. and Polkovnikov, A.},
  editor = {Chandra, Anjan Kumar and Das, Arnab and Chakrabarti, Bikas K.},
  year = 2010,
  volume = {802},
  pages = {75--114},
  publisher = {Springer Berlin Heidelberg},
  address = {Berlin, Heidelberg},
  doi = {10.1007/978-3-642-11470-0_4},
  urldate = {2026-02-15},
  isbn = {978-3-642-11469-4 978-3-642-11470-0},
  langid = {english}
}

@article{dziarmagaDynamicsQuantumPhase2010,
  title = {Dynamics of a Quantum Phase Transition and Relaxation to a Steady State},
  author = {Dziarmaga, Jacek},
  year = 2010,
  month = nov,
  journal = {Advances in Physics},
  volume = {59},
  number = {6},
  pages = {1063--1189},
  issn = {0001-8732, 1460-6976},
  doi = {10.1080/00018732.2010.514702},
  urldate = {2026-02-18},
  langid = {english}
}

@article{espositoNonequilibriumFluctuationsFluctuation2009,
  title = {Nonequilibrium Fluctuations, Fluctuation Theorems, and Counting Statistics in Quantum Systems},
  author = {Esposito, Massimiliano and Harbola, Upendra and Mukamel, Shaul},
  year = 2009,
  month = dec,
  journal = {Reviews of Modern Physics},
  volume = {81},
  number = {4},
  pages = {1665--1702},
  issn = {0034-6861, 1539-0756},
  doi = {10.1103/RevModPhys.81.1665},
  urldate = {2026-02-18},
  copyright = {http://link.aps.org/licenses/aps-default-license},
  langid = {english}
}

@article{feiUniversalScalingWork2021,
  title = {Universal Scaling of Work Statistics in Conformal Field Theory Models},
  author = {Fei, Zhaoyu and Sun, C. P.},
  year = 2021,
  month = apr,
  journal = {Physical Review B},
  volume = {103},
  number = {14},
  pages = {144204},
  issn = {2469-9950, 2469-9969},
  doi = {10.1103/PhysRevB.103.144204},
  urldate = {2026-02-14},
  langid = {english}
}

@article{feiWorkStatisticsQuantum2020,
  title = {Work {{Statistics}} across a {{Quantum Phase Transition}}},
  author = {Fei, Zhaoyu and Freitas, Nahuel and Cavina, Vasco and Quan, H. T. and Esposito, Massimiliano},
  year = 2020,
  month = may,
  journal = {Physical Review Letters},
  volume = {124},
  number = {17},
  pages = {170603},
  issn = {0031-9007, 1079-7114},
  doi = {10.1103/PhysRevLett.124.170603},
  urldate = {2026-02-14},
  langid = {english}
}

@article{funoPathIntegralApproach2018,
  title = {Path {{Integral Approach}} to {{Quantum Thermodynamics}}},
  author = {Funo, Ken and Quan, H. T.},
  year = 2018,
  month = jul,
  journal = {Physical Review Letters},
  volume = {121},
  number = {4},
  pages = {040602},
  issn = {0031-9007, 1079-7114},
  doi = {10.1103/PhysRevLett.121.040602},
  urldate = {2026-02-14},
  langid = {english}
}

@article{gelfandIntegrationFunctionalSpaces1960,
  title = {Integration in {{Functional Spaces}} and Its {{Applications}} in {{Quantum Physics}}},
  author = {Gel'fand, I. M. and Yaglom, A. M.},
  year = 1960,
  month = jan,
  journal = {Journal of Mathematical Physics},
  volume = {1},
  number = {1},
  pages = {48--69},
  issn = {0022-2488, 1089-7658},
  doi = {10.1063/1.1703636},
  urldate = {2026-05-09},
  langid = {english}
}

@article{husimiMiscellaneaElementaryQuantum1953,
  title = {Miscellanea in {{Elementary Quantum Mechanics}}, {{II}}},
  author = {Husimi, K.},
  year = 1953,
  month = apr,
  journal = {Progress of Theoretical Physics},
  volume = {9},
  number = {4},
  pages = {381--402},
  issn = {0033-068X, 1347-4081},
  doi = {10.1143/ptp/9.4.381},
  urldate = {2026-05-25},
  langid = {english}
}

@article{jarzynskiNonequilibriumEqualityFree1997,
  title = {Nonequilibrium {{Equality}} for {{Free Energy Differences}}},
  author = {Jarzynski, C.},
  year = 1997,
  month = apr,
  journal = {Physical Review Letters},
  volume = {78},
  number = {14},
  pages = {2690--2693},
  issn = {0031-9007, 1079-7114},
  doi = {10.1103/PhysRevLett.78.2690},
  urldate = {2026-05-25},
  copyright = {http://link.aps.org/licenses/aps-default-license},
  langid = {english}
}

@article{Kibble1976,
  title = {Topology of Cosmic Domains and Strings},
  author = {Kibble, T W B},
  year = 1976,
  month = aug,
  journal = {Journal of Physics A: Mathematical and General},
  volume = {9},
  number = {8},
  pages = {1387--1398},
  issn = {0305-4470, 1361-6447},
  doi = {10.1088/0305-4470/9/8/029},
  urldate = {2026-02-16},
  langid = {english}
}

@book{kleinertPathIntegralsQuantum2004,
  title = {Path {{Integrals}} in {{Quantum Mechanics}}, {{Statistics}}, {{Polymer Physics}}, and {{Financial Markets}}},
  author = {Kleinert, Hagen},
  year = 2004,
  month = mar,
  edition = {3},
  publisher = {WORLD SCIENTIFIC},
  doi = {10.1142/5057},
  urldate = {2026-05-09},
  isbn = {978-981-238-106-4 978-981-256-219-7},
  langid = {english}
}

@article{lewisExactQuantumTheory1969,
  title = {An {{Exact Quantum Theory}} of the {{Time-Dependent Harmonic Oscillator}} and of a {{Charged Particle}} in a {{Time-Dependent Electromagnetic Field}}},
  author = {Lewis, H. R. and Riesenfeld, W. B.},
  year = 1969,
  month = aug,
  journal = {Journal of Mathematical Physics},
  volume = {10},
  number = {8},
  pages = {1458--1473},
  issn = {0022-2488, 1089-7658},
  doi = {10.1063/1.1664991},
  urldate = {2026-05-29},
  langid = {english}
}

@article{polkovnikovColloquiumNonequilibriumDynamics2011,
  title = {{\emph{Colloquium}} : {{Nonequilibrium}} Dynamics of Closed Interacting Quantum Systems},
  shorttitle = {{\emph{Colloquium}}},
  author = {Polkovnikov, Anatoli and Sengupta, Krishnendu and Silva, Alessandro and Vengalattore, Mukund},
  year = 2011,
  month = aug,
  journal = {Reviews of Modern Physics},
  volume = {83},
  number = {3},
  pages = {863--883},
  issn = {0034-6861, 1539-0756},
  doi = {10.1103/RevModPhys.83.863},
  urldate = {2026-02-15},
  copyright = {http://link.aps.org/licenses/aps-default-license},
  langid = {english}
}

@article{polkovnikovUniversalAdiabaticDynamics2005,
  title = {Universal Adiabatic Dynamics in the Vicinity of a Quantum Critical Point},
  author = {Polkovnikov, Anatoli},
  year = 2005,
  month = oct,
  journal = {Physical Review B},
  volume = {72},
  number = {16},
  pages = {161201},
  issn = {1098-0121, 1550-235X},
  doi = {10.1103/PhysRevB.72.161201},
  urldate = {2026-02-18},
  copyright = {http://link.aps.org/licenses/aps-default-license},
  langid = {english}
}

@article{RevModPhys.20.367,
  title = {Space-Time Approach to Non-Relativistic Quantum Mechanics},
  author = {Feynman, R. P.},
  journal = {Rev. Mod. Phys.},
  volume = {20},
  number = {2},
  pages = {367--387},
  numpages = {0},
  year = {1948},
  month = {Apr},
  publisher = {American Physical Society},
  doi = {10.1103/RevModPhys.20.367},
  url = {https://link.aps.org/doi/10.1103/RevModPhys.20.367}
}

@article{senDefectProductionNonlinear2008,
  title = {Defect {{Production}} in {{Nonlinear Quench}} across a {{Quantum Critical Point}}},
  author = {Sen, Diptiman and Sengupta, K. and Mondal, Shreyoshi},
  year = 2008,
  month = jul,
  journal = {Physical Review Letters},
  volume = {101},
  number = {1},
  pages = {016806},
  issn = {0031-9007, 1079-7114},
  doi = {10.1103/PhysRevLett.101.016806},
  urldate = {2026-02-18},
  copyright = {http://link.aps.org/licenses/aps-default-license},
  langid = {english}
}

@article{talknerFluctuationTheoremsWork2007,
  title = {Fluctuation Theorems: {{Work}} Is Not an Observable},
  shorttitle = {Fluctuation Theorems},
  author = {Talkner, Peter and Lutz, Eric and H{\"a}nggi, Peter},
  year = 2007,
  month = may,
  journal = {Physical Review E},
  volume = {75},
  number = {5},
  pages = {050102},
  issn = {1539-3755, 1550-2376},
  doi = {10.1103/PhysRevE.75.050102},
  urldate = {2026-05-05},
  copyright = {http://link.aps.org/licenses/aps-default-license},
  langid = {english}
}

@article{zhangWorkStatisticsQuantum2022,
  title = {Work Statistics across a Quantum Critical Surface},
  author = {Zhang, Fan and Quan, H. T.},
  year = 2022,
  month = feb,
  journal = {Physical Review E},
  volume = {105},
  number = {2},
  pages = {024101},
  issn = {2470-0045, 2470-0053},
  doi = {10.1103/PhysRevE.105.024101},
  urldate = {2026-02-19},
  langid = {english}
}

@article{Zurek1985,
  title = {Cosmological Experiments in Superfluid Helium?},
  author = {Zurek, Wojciech H.},
  year = 1985,
  journal = {Nature},
  volume = {317},
  number = {6037},
  pages = {505--508},
  publisher = {Nature Publishing Group UK London},
  urldate = {2026-02-17}
}

@article{zurekDynamicsQuantumPhase2005,
  title = {Dynamics of a {{Quantum Phase Transition}}},
  author = {Zurek, Wojciech H. and Dorner, Uwe and Zoller, Peter},
  year = 2005,
  month = sep,
  journal = {Physical Review Letters},
  volume = {95},
  number = {10},
  pages = {105701},
  issn = {0031-9007, 1079-7114},
  doi = {10.1103/PhysRevLett.95.105701},
  urldate = {2026-02-17},
  copyright = {http://link.aps.org/licenses/aps-default-license},
  langid = {english}
}

@article{qiuPathIntegralApproach2020,
  author = {Qiu, Tian and Fei, Zhaoyu and Pan, Rui and Quan, H. T.},
  title = {Path-Integral Approach to the Calculation of the Characteristic Function of Work},
  journal = {Physical Review E},
  volume = {101},
  number = {3},
  pages = {032111},
  year = {2020},
  month = mar,
  doi = {10.1103/PhysRevE.101.032111},
  eprint = {1908.09731},
  archiveprefix = {arXiv},
  primaryclass = {cond-mat.stat-mech}
}

@misc{qiao2026field,
  title = {A Field-Theoretic Framework for Work Statistics and Universal Scaling in Non-equilibrium Phase Transitions},
  author = {Qiao, Yanbo and Xu, Ruohan and Quan, H. T.},
  year = {2026},
  eprint = {2606.30503},
  archiveprefix = {arXiv},
  url = {https://arxiv.org/abs/2606.30503}
}

@article{dziarmagaQuantumIsingExact2005,
  title = {Dynamics of a Quantum Phase Transition: Exact Solution of the Quantum Ising Model},
  author = {Dziarmaga, Jacek},
  journal = {Physical Review Letters},
  volume = {95},
  pages = {245701},
  year = {2005},
  doi = {10.1103/PhysRevLett.95.245701}
}

@article{kolodrubetzClassifyingMeasuringGeometry2013,
  author = {Kolodrubetz, Michael and Gritsev, Vladimir and Polkovnikov, Anatoli},
  title = {Classifying and measuring geometry of a quantum ground state manifold},
  journal = {Physical Review B},
  volume = {88},
  pages = {064304},
  year = {2013},
  doi = {10.1103/PhysRevB.88.064304},
  eprint = {1305.0568},
  archivePrefix = {arXiv}
}

\end{document}